\documentclass[conference]{IEEEtran}
\usepackage{graphicx}
\usepackage{cite}
\usepackage{amsmath}
\usepackage{amsfonts}
\usepackage{amssymb}
\usepackage{epsfig}

\newcommand{\bm}{\mathbf}
\newcommand{\be}{\begin{equation}}
\newcommand{\ee}{\end{equation}}
\newcommand{\bse}{\begin{subequations}}
\newcommand{\ese}{\end{subequations}}
\newcommand{\bea}{\begin{eqnarray}}
\newcommand{\eea}{\end{eqnarray}}
\newcommand{\x}{{\bm x}}

\newcommand{\ba}{{\bm a}}

\newcommand{\br}{{\bm r}}

\newcommand{\f}{{\bm f}}

\newcommand{\bA}{{\bm A}}

\newcommand{\bI}{{\bm I}}
\newcommand{\bR}{{\bm R}}
\newcommand{\bW}{{\bm W}}

\newcommand{\bF}{{\bf F}}
\newcommand{\bD}{{\bf D}}

\newcommand{\bS}{{\bf S}}
\newcommand{\bG}{{\bf G}}
\newcommand{\bH}{{\bf H}}
\newcommand{\bX}{{\bf X}}

\newcommand{\bs}{{\bf s}}

\newcommand{\brho}{\mbox{\boldmath{$\rho$}}}
\newcommand{\expect}{{\mathbb{E}}}
\IEEEoverridecommandlockouts
\begin{document}

\raggedbottom
%

\title{OTFS Without CP in Massive MIMO: Breaking Doppler Limitations with TR-MRC and Windowing\\\thanks{This publication has emanated from research conducted with the financial support of Science Foundation Ireland under Grant number 19/FFP/7005.}}

\author{\normalsize Danilo Lelin Li, Arman Farhang
\\Electronic \& Electrical Engineering Department, Trinity College Dublin, Ireland \\
Email: lelinlid@tcd.ie, Arman.Farhang@tcd.ie.
\vspace{-0.2cm}
}

\maketitle

%
%
\begin{abstract}
Orthogonal time frequency space (OTFS) modulation has recently emerged as an effective waveform to tackle the linear time-varying channels. In OTFS literature, approximately constant channel gains for every group of samples within each OTFS block are assumed. This leads to limitations for OTFS on the maximum Doppler frequency that it can tolerate. Additionally, presence of cyclic prefix (CP) in OTFS signal limits the flexibility in adjusting its parameters to improve its robustness against channel time variations. Therefore, in this paper, we study the possibility of removing the CP overhead from OTFS and breaking its Doppler limitations through multiple antenna processing in the large antenna regime. We asymptotically analyze the performance of time-reversal maximum ratio combining (TR-MRC) for OTFS without CP. We show that doubly dispersive channel effects average out in the large antenna regime when the maximum Doppler shift is within OTFS limitations. However, for considerably large Doppler shifts exceeding OTFS limitations, a residual Doppler effect remains. Our asymptotic derivations reveal that this effect converges to scaling of the received symbols in delay dimension with the samples of a Bessel function that depends on the maximum Doppler shift. Hence, we propose a novel residual Doppler correction (RDC) windowing technique that can break the Doppler limitations of OTFS and lead to a performance close to that of the linear time-invariant channels. Finally, we confirm the validity of our claims through simulations.
\end{abstract}

%
%

\section{Introduction}
The new generation of services and applications in 6G mobile systems present many challenging requirements, including extremely low latency and ultra-high reliability of the wireless links. These challenges are more noticeable in mission-critical applications such as autonomous driving that require safe and rapid reactions and cannot tolerate the wireless link becoming unreliable. Loss of reliability can be due to the fast time variations of the channel caused by mobility. Dealing with time-varying channels has a long history and rigorous foundations in the development of time-frequency domain signaling schemes, \cite{OFDM}. However, conventional solutions require fast channel tracking and/or large signaling overheads leading to significant latency issues.

As one of the key building blocks that will underpin next generation networks, the air interface needs to be highly resilient to inherent wireless channel-fading effects, i.e. multipath and Doppler effects. While orthogonal frequency division multiplexing (OFDM) has been the technology of the choice for 4G and 5G systems, its high sensitivity to Doppler effects due to the time variations of the channel motivate the need for a more robust waveform. Thus, a new waveform called orthogonal time frequency space (OTFS) has recently emerged, \cite{OTFS}. OTFS has a novel and fresh approach to waveform design as it deploys delay-Doppler instead of time-frequency domain for data transmission. An important aspect of placing data symbols in delay-Doppler domain is that the response in this domain relating the input to the output of the channel is sparse and time-invariant, \cite{OTFS,OTFS2,mmwave}. Hence, OTFS requires much lower training overheads and much slower channel tracking requirements than OFDM. OTFS, in its original form \cite{OTFS}, can be implemented on top of OFDM and its signal can be formed by concatenating multiple OFDM symbols in time. However, OFDM may suffer from bandwidth efficiency loss due to its redundant cyclic prefix (CP), whose length can be up to $25\%$ of the symbol duration in its extended form as devised in the recent 3GPP 5G NR standard \cite{lte2009evolved2}.

An important point to note is that the CP overhead makes the symbols longer and hence the time variation of the channel from one symbol to the next is increased. Additionally, the presence of CP can become a bottleneck when shortening the OFDM symbols within each OTFS block to tackle the time variations of the channel. Therefore, the authors in \cite{PracticalPulse} have shown that the CP overhead can be reduced to only one CP at the beginning of each OTFS block instead of having multiple CPs within one block of OTFS. Recently, the authors in \cite{OFDMwoCP} have shown that in massive multiple input multiple output (MIMO) systems one can dispense with the redundant CP of OFDM by application of time-reversal maximum ratio combining (TR-MRC) in linear time invariant (LTI) channels. In a more recent work, the authors in \cite{TRsinglecarrier} extended the results of \cite{OFDMwoCP} to single carrier transmission in LTI channels. Massive MIMO is one of the key 5G technologies that can average out the effects of noise and interference by deploying a large number of antennas at the base station (BS). Motivated by the results in \cite{OFDMwoCP,TRsinglecarrier}, in this paper, we investigate application of TR-MRC to OTFS when the CP overhead is completely removed in linear time varying (LTV) channels and address the following research question. 

\textit{``Can TR-MRC average out doubly spread channel effects and break the Doppler limitations of OTFS?"}

There are a number of works emerging in recent literature on OTFS addressing multiple fronts of high Doppler scenarios, \cite{ramachandran2018mimo,8377159,8727425,NewPathMMIMO,MMIMODownlink}. One point that all these works have in common is the Doppler limitation due to the requirement of a near constant channel over each OFDM symbol within the OTFS blocks. Furthermore, all these works consider either one CP per OFDM symbol, i.e., multiple CPs in each OTFS block, or one CP per OTFS block. To resolve these issues, in this paper, we apply TR-MRC to OTFS without CP. With the assumption of locally time invariant channels over each OFDM symbol in the OTFS blocks, we show that both delay and Doppler spread of the channel at the time-reversal combiner output average out as the number of BS antennas grows large. However, for considerably large Doppler shifts, this assumption does not hold and a residual Doppler effect remains after time-reversal combining. An interesting finding of this paper is that in the asymptotic regime, this effect reduces to scaling of the received symbols in the delay dimension with the samples of a Bessel function that depends on the maximum Doppler frequency. Based on this observation, we propose a novel solution to the Doppler limitation of OTFS through application of a residual Doppler correction (RDC) window. 

To corroborate our claims in this paper, we numerically analyze the performance of our proposed technique when the CP overhead is completely removed from OTFS in terms of both signal-to-interference-plus-noise ratio (SINR) and bit error rate (BER) performance. Based on our simulations, when the maximum Doppler shift is within the OTFS limitations, the SINR of our proposed technique is a linear function of the number of BS antennas. This confirms the efficacy of TR-MRC in averaging out both delay and Doppler spread of the LTV channel. However, when the maximum Doppler shift exceeds OTFS limitations, the SINR performance is a linear function of the number of BS antennas only with application of our proposed RDC windowing method, whereas otherwise the SINR saturates. Furthermore, thanks to the absence of CP, we show that the OFDM symbols within each OTFS block can be shortened and lead to a higher resilience against Doppler effect. To conclude our analysis, we show that our proposed RDC windowing technique leads to a BER performance close to that of LTI channels even for substantially large values of Doppler spread.

The rest of the paper is organized as follows. In Section~\ref{sec:sysmod}, we explain the system model for OFDM-based OTFS without CP in massive MIMO. Section~\ref{sec:TRinTD} presents our proposed TR-MRC with RDC windowing technique. In Section~\ref{sec:numres}, we confirm the efficacy of our proposed technique by simulations. Finally, we conclude the paper in Section \ref{sec:conc}.

{\it Notations}: Matrices, vectors and scalar quantities are denoted by boldface uppercase, boldface lowercase and normal
letters, respectively. $[\bA]_{m,n}$
represents the element in the $m^{\rm{th}}$ row and $n^{\rm{th}}$ column of $\bA$
. $\bI_M$ and $\textbf{0}_{M\times N}$ are the identity and zero matrices of the sizes $M \times M$ and $M \times N$, respectively. $\bD= {\rm diag}\{ \ba \}$ is a diagonal matrix with the diagonal elements  formed by $\ba$. The superscripts $(\cdot)^{\rm T}$ and $(\cdot)^{*}$ indicate transpose and conjugate operations, respectively. $|\cdot |$ and $\expect\{\cdot \}$ are the absolute value and expected value operators, respectively. $\delta(\cdot)$ is the Dirac delta function. Finally, $\bF_M$ is the normalized $M$-point discrete Fourier transform (DFT) matrix with the elements $[\bF_M]_{mn}$ = $\frac{1}{\sqrt{M}}e^{\frac{-j2\pi mn}{M}}$, for $m,~n = 0,..., M-1$ and $\f_{M,m}$ the $m^{\rm th}$ column of $\bF_M$. 

\vspace{-0.1cm}
\section{System Model}
\label{sec:sysmod}

\begin{figure*}[!t]
\normalsize

   {\footnotesize \begin{align}
   \label{eq:matsall}
    \bH^{(n,n-1)}_{n,q}&=
    \begin{pmatrix}
        0  \!\!\!\!\!& \dots \!\!\!\!\!& {{{h}}}_{q} [ L-1 , Mn ] \!\!\!\!\!& \dots \!\!\!\!\!& {{{h}}}_{q} [ 1 , Mn ]\\
        \vdots \!\!\!\!\!& \ddots \!\!\!\!\! & \ddots \!\!\!\!\!& \ddots \!\!\!\!\!& \vdots\\
        0  \!\!\!\!\!& \ddots \!\!\!\!\!& \ddots \!\!\!\!\!& \ddots \!\!\!\!\!& {{{h}}}_{q} [ L-1 , Mn +L-2 ]\\
        \vdots \!\!\!\!\! & \ddots \!\!\!\!\! & \ddots \!\!\!\!\!& \ddots \!\!\!\!\!& \vdots\\
        0  \!\!\!\!\!& \dots \!\!\!\!\! & \dots \!\!\!\!\!& \dots\!\!\!\!\!&0\\
    \end{pmatrix},
    \bH^{(n,n)}_{n,q} =
    \begin{pmatrix}
        {{{h}}}_{q} [ 0, Mn ]   \!\!\!\!\!& \dots \!\!\!\!\!& 0\\
        \vdots \!\!\!\!\!& \ddots   \!\!\!\!\!& \vdots\\
        {{{h}}}_{q} [ L-1 , Mn +L-1 ]  \!\!\!\!\!& \ddots \!\!\!\!\!& 0\\
        \vdots  \!\!\!\!\!& \ddots  \!\!\!\!\!& \vdots\\
        0   \!\!\!\!\!& \ddots \!\!\!\!\!& {{{h}}}_{q} [ 0, M(n+1) -1 ]\\
        \vdots \!\!\!\!\!& \ddots  \!\!\!\!\!& \vdots\\
        0   \!\!\!\!\!& \dots \!\!\!\!\!& {{{h}}}_{q} [ L-1, M(n+1) +L-2 ]\\
    \end{pmatrix},\nonumber\\
      \bH^{(n,n+1)}_{n,q}&=
 \begin{pmatrix}
        0 \!\!\!\!\!& \dots  \!\!\!\!\!   & \dots \!\!\!\!\!& \dots\!\!\!\!\! & 0\\
        \vdots\!\!\!\!\! & \ddots    \!\!\!\!\!& \ddots \!\!\!\!\!& \ddots \!\!\!\!\!& \vdots\\
        {{{h}}}_{q} [ 0 , M(n+1) ] \!\!\!\!\! & \ddots \!\!\!\!\!   & \ddots\!\!\!\!\! & \ddots \!\!\!\!\!& 0\\
        \vdots \!\!\!\!\!& \ddots \!\!\!\!\!    & \ddots\!\!\!\!\! & \ddots \!\!\!\!\!& \vdots\\
        {{{h}}}_{q} [ L-2 ,M(n+1) +L-2 ] \!\!\!\!\!& \dots  \!\!\!\!\! & {{{h}}}_{q} [ 0 , M(n+1) +L-2 ]\!\!\!\!\! & \dots \!\!\!\!\!& 0\\
    \end{pmatrix}.\nonumber
\end{align}}
\vspace{-0.2cm}
\hrulefill
\vspace*{4pt}
\end{figure*}

We consider an OFDM-based OTFS system in a large scale MIMO setup where a single antenna user is communicating with a BS that is equipped with $Q$ antennas. To better cope with the time variations of the channel, we shorten the OFDM symbols by removing the CP overhead. Let the $M\times N$ matrix $\bX$ contain the quadrature amplitude modulated (QAM) data symbols in the delay-Doppler domain. The elements of $\bX$ are independent and identically distributed (i.i.d.) zero-mean complex random variables of unit variance. In the first stage of OTFS modulation, the data symbols are first mapped to the time-frequency plane with $N$ time slots and $M$ frequency bins by an inverse symplectic finite Fourier transform (ISFFT) operation, \cite{OTFS}. This can be implemented by performing $M$-point DFT and $N$-point  inverse DFT (IDFT) operations along the columns and rows of $\bX$, respectively, i.e., $\bF_M\bX\bF_N^{\rm H}$. In this paper, we consider OTFS with rectangular transmit and receive window functions. In the second stage, the resulting time-frequency samples are fed into an OFDM modulator and the OTFS transmit signal in the absence of CP is formed as $\bS= \bF^{\rm H}_M (\bF_M \bX \bF^{\rm H}_N)=\bX \bF^{\rm H}_N$. Hence, the OFDM-based OTFS modulation reduces to $N$-point IDFT operations along the rows of $\bX$, \cite{farhang2017low}. After parallel-to-serial conversion of $\bS$, the OTFS transmit signal can be formed as $\bs= [\bs_0^{\rm T},\ldots,\bs_{N-1}^{\rm T}]^{\rm T}$ where $\bs_n$ indicates the $n^{\rm th}$ column of $\bS$, i.e., 
\begin{equation}
    \label{eq:sn}
        {\bs}_n ={\bX} \f^*_{N,n}
        =\frac{1}{\sqrt{N}}\sum^{N-1}_{i=0} {\x}_i e^{\frac{j2\pi n i}{N}},
\end{equation}
and ${\x}_i$ denotes the $i^{\rm th}$ column of $\bX$.

After the digital-to-analog conversion of the signal $\bs$, the base band continuous time signal, $s(t)$ goes through the LTV channel. Considering the same signal-to-noise ratio (SNR) and statistically independent channels between the mobile terminal and the BS antennas, the received signal at the BS antenna $q$ can be written as 
\begin{equation}\label{eq:chconv}
    r_q(t)=\int \int h_q(\tau,\upsilon) s(t-\tau)e^{j2\pi \upsilon (t-\tau)} d\upsilon d\tau + \eta_q(t), 
\end{equation}
where $\eta_q(t)$ is the additive white Gaussian noise and $h_q(\tau,\upsilon)=\sum_{p=0}^{P-1} \alpha_{p,q} \delta (\tau - \tau_{p,q}) \delta (\upsilon - \upsilon_{p,q})$ is the sparse channel response with $P$ paths in delay-Doppler domain between the user and BS antenna $q$. 
The parameters $\alpha_{p,q}$, $\tau_{p,q}$ and $\upsilon_{p,q}$ represent the path gains, delays and Doppler shifts of path $p$ at antenna $q$, respectively. We assume the same power delay profile (PDP) for the channels between the user and all the BS antennas and the path gains at different antennas to be i.i.d. complex Gaussian random variables with zero mean and variance of $\rho (p)$, i.e., $\alpha_{p,q} \sim \mathcal{CN}(0,\rho (p)),\forall q$. We consider a normalized PDP, $\brho =[\rho (0),...,\rho (P-1)]^{\rm{T}}$ where $\sum_{p=0}^{P-1} \rho (p)=1$. 
In practical systems, the sampling period is short enough to approximate the path delays to the nearest sampling points. Consequently, we do not consider the fractional delays and each $\tau_{p,q}$ is approximated as an integer multiple of the sampling period $T_{\rm s}$, i.e., $\tau_{p,q}\approx \ell_{\tau_{p,q}} T_{\rm s}$ where $\ell_{\tau_{p,q}} \in [0,L-1]$ and $L$ is the number of delay taps after discretization. Thus, the received signal at antenna $q$ after analog-to-digital conversion with the sampling period $T_{\rm s}$, i.e., $ r_q( \ell T_{\rm s} ) = r_q[\ell]$, can be expressed as
\begin{align}
    \begin{split}\label{eq:linch}
        r_q[\ell] 
        =\sum_{k=0}^{L-1}  {{{h}}}_{q} [k,\ell] s[\ell-k]+ \eta_q[\ell],
    \end{split}
\end{align}
where ${{{h}}}_{q} [k,\ell] = \sum_{p=0}^{P-1} \alpha_{p,q} e^{j2\pi \upsilon_{p,q} (\ell-k) T_{\rm{s}}} \delta [k-\ell_{\tau_{p,q}}]$ is the channel response in delay-time domain and ${{{h}}}_{q} [k,\ell]=0$ for any $k \notin [0,L-1]$.

After serial-to-parallel conversion of each received OTFS block of size $MN$ samples at a given antenna $q$, we form an $M \times N$ matrix $\bR_q=[{\br}_{0,q},\ldots,{\br}_{N-1,q}]$ with the columns ${\br}_{n,q}=[r_q[nM],\ldots,r_q[(n+1)M-1]]^{\rm T}$. In the first stage of OTFS demodulation, the columns of $\bR_q$ are fed into an OFDM demodulator to obtain the received signal in time-frequency domain as $\bF_M \bR_q$. The resulting signal samples are then translated to the delay-Doppler domain at the second OTFS demodulation stage through a SFFT operation, i.e., $\hat{\bX}_q = \bF_M^{\rm H}(\bF_M \bR_q)\bF_N=\bR_q\bF_N$ where $\hat{\bX}_q=[{\hat{\x}}_{0,q},...,{\hat{\x}}_{N-1,q}]$ and
\be
    \label{eq:tildaxn1}
        \hat{\x}_{n,q} ={\bR_q} \f_{N,n}
        =\frac{1}{\sqrt{N}}\sum^{N-1}_{i=0}{\br}_{i,q} e^{\frac{-j2\pi n i}{N}}.
\ee

\vspace{1mm}
\section{Proposed TR-MRC with RDC Windowing}\label{sec:TRinTD}
Complete removal of CP shortens OFDM symbols and brings a higher resilience to time variations of the channel. However, this leads to intersymbol interference (ISI) and inter-block interference (IBI) between the adjacent OFDM symbols and OTFS blocks, respectively. Previous studies, \cite{OFDMwoCP,TRsinglecarrier}, have shown that TR-MRC can mitigate the ISI and ICI effects for OFDM and single carrier transmissions in massive MIMO systems. 
Hence, in this section, we apply TR-MRC to OFDM-based OTFS without CP. We show that the interference caused by both delay spread, i.e., ISI and IBI, and time variations of the channel can be effectively alleviated as the number of BS antennas grows large. 

Considering the received OFDM symbol $n$ at antenna $q$ with the samples due to the transient of the channel from (\ref{eq:linch}), we can form the $(M+L-1)\times 1$ vector $\tilde{\br}_{n,q}=[r_q[nM],\ldots,r_q[(n+1)M+L-2]]^{\rm T}$ as
\begin{equation}\label{eq:channel}
    \Tilde{\br}_{n,q}=\bH^{(n,n-1)}_{n,q} \bs_{n-1}+ \bH^{(n,n)}_{n,q} \bs_n+\bH^{(n,n+1)}_{n,q} \bs_{n+1}+\boldsymbol{\eta}_{n,q}.
\end{equation} 
The $(M+L-1)\times M$ convolution matrices $\bH^{(n,n)}_{n,q}$, $\bH^{(n,n-1)}_{n,q}$ and $\bH^{(n,n+1)}_{n,q}$ when multiplied to the symbols $\bs_n$, $\bs_{n-1}$ and $\bs_{n+1}$ form the channel affected symbol $n$, the tail of the  symbol $n-1$ that overlaps with the first $L-1$ samples of the symbol $n$, and the first $L-1$ samples of the symbol $n+1$ that overlap with the tail of the symbol $n$, respectively. The elements of these matrices are represented as $[\bH^{(n,n)}_{n,q}]_{ab}={{{h}}}_{q} [ a-b, Mn +a]$, $[\bH^{(n,n-1)}_{n,q}]_{ab}={{{h}}}_{q} [ M-b+a ,Mn +a]$ and $[\bH^{(n,n+1)}_{n,q}]_{ab}={{{h}}}_{q} [ a-b-M ,Mn +a]$, where $a=0,\ldots,M+L-2$ and $b=0,\ldots,M-1$. It is worth mentioning that these matrices follow a Toeplitz-like structure (but, not quite Toeplitz because of the time variation of the channel), shown on the top of this page, where each diagonal represents a channel tap that varies every sample according to its respective Doppler shift. 
As mentioned earlier, the absence of CP leads to IBI between OTFS blocks apart from ISI. This effect is represented using the same matrices as the ones that are shown in (\ref{eq:channel}). 
Hence, we treat IBI in the same way as ISI. 

In the following subsections, we asymptotically analyze OTFS without CP using TR-MRC as the number of BS antennas tends to infinity for both LTI and LTV channels. 
Based on our derivations and analysis, we propose a novel windowing technique tailored to the channel correlation function to tackle the residual Doppler effect due to the time variations of the channel within each OFDM symbol.

\vspace{3mm}
\subsection{TR-MRC for OTFS in LTI channels}\label{subsec:LTI}
For an LTI channel, there is no time variation over the OTFS blocks, i.e., ${{{h}}}_{q} [ k, 0] = \ldots ={{{h}}}_{q} [ k, MN]$, $\forall k$. 
Thus, the convolution matrices on the top of the previous page become Toeplitz matrices.
In TR-MRC, the received signals on the BS antennas $q$ for OFDM time symbol $n$ are first pre-filtered with the corresponding time-reversed and conjugated CIRs. Then the resulting signals at different antennas are combined and the first $L-1$ samples at the filter output, i.e., the time-reversal filter transient, are discarded by the rectangular window $\bW=[\textbf{0}_{M\times(L-1)}, \bI_M]$. Thus, we have 
\be
\vspace{1mm}
    \label{eq:rTR}
        \br^{\rm{TR}}_{n}=\frac{1}{Q}\sum_{q=0}^{Q-1}\bW \bH^{\rm{TR}}_{n,q} \Tilde{\br}_{n,q},
\ee
where the time-reversal filter for antenna $q$ can be represented by $\bH^{\rm{TR}}_{n,q}$ with the elements $[\bH^{\rm{TR}}_{n,q}]_{ab}={{{h}}}_{q}^* [ L-1-(a-b), 0]$ that are independent of $n$, for $a,b=0,...,M+L-2$.

Substituting $\Tilde{\br}_{n,q}$ from (\ref{eq:channel}) into (\ref{eq:rTR}), we have
\be
\vspace{1mm}
    \label{eq:LTIrtr}
      {\br}_{n}^{\rm{TR}}= \bG^{(n,n-1)}_{n} \bs_{n-1}+ \bG^{(n,n)}_{n} \bs_n + \bG^{(n,n+1)}_{n} \bs_{n+1}+\boldsymbol{\eta}_{n}^{\prime},
      \vspace{1mm}
\ee
where $\!\!\bG^{(n,n-1)}_{n}\!\!\!\!=\!\!\!\frac{1}{Q}\!\sum_{q=0}^{Q-1}\!\bW \bH^{\rm{TR}}_{n,q} \bH^{(n,n-1)}_{n,q}\!\!$, $\bG^{(n,n)}_{n}\!\!\!\!=\!\!\!\! \frac{1}{Q}\!\!\sum_{q=0}^{Q-1}$ $ \bW \bH^{\rm{TR}}_{n,q} \bH^{(n,n)}_{n,q}$, $\bG^{(n,n+1)}_{n} =  \frac{1}{Q}\sum_{q=0}^{Q-1} \bW \bH^{\rm{TR}}_{n,q} \bH^{(n,n+1)}_{n,q}$ and $\boldsymbol{\nu}_{n}^{\prime} =\frac{1}{Q}\sum_{q=0}^{Q-1} \bW \bH^{\rm{TR}}_{n,q} \boldsymbol{\nu}_{n,q}$. The elements of the matrices are given by $[\bG^{(n,n-1)}_{n}]_{a,b}=g[M+a-b,0]$, $[\bG^{(n,n)}_{n}]_{a,b}=g[a-b,0]$ and $[\bG^{(n,n+1)}_{n}]_{a,b}=g[a-b-M,0]$ for $a,b=0,\ldots,M-1$, and $g[i,0]=\frac{1}{Q}\sum_{q=0}^{Q-1} \sum_{k=0}^{L-1} {{{h}}}_q[k,0] {{{h}}}^*_q[k-i,0]$ represents the channel correlation function after combining.

As the number of BS antennas $Q$ tends to infinity, due to the law of large numbers, the equivalent time-reversal combining matrices in (\ref{eq:LTIrtr}) converge almost surely to the respective expected value.
Hence, as $Q$ tends to infinity, $\bG^{(n,n)}_{n}$ tends to an identity matrix while $\bG^{(n,n-1)}_{n}$ and $\bG^{(n,n+1)}_{n}$ tend to zero. This is due to the fact that the elements on the main diagonal of $\bG^{(n,n)}_{n}$ tend to $\sum^{L-1}_{k=0} \expect \{ {{{h}}}_q[k,0] {{{h}}}^*_q[k,0] \}=\sum^{P-1}_{p=0}\expect \{ | \alpha_{p,q}  |^2 \}=1$, while the off-diagonal elements of $\bG^{(n,n)}_{n}$ tend to sum of uncorrelated terms of the form $\expect \{ {{{h}}}_q[k,0] {{{h}}}^*_q[j,0] \}=\expect \{ \alpha_{p,q} \alpha^*_{p',q} \} = 0$, when $k \neq j$ and consequently $p \neq p'$.

Finally, TR-MRC output signal is fed into the OTFS demodulator to retrieve the delay-Doppler data symbols,
\be\label{eq:xtrLTI}
\vspace{1mm}
    \hat{\x}_{n}^{\rm{TR}}=\frac{1}{\sqrt{N}}\sum^{N-1}_{i=0}{\br}^{\rm{TR}}_{i} e^{\frac{-j2\pi n i}{N}}.
\ee
Based on the above discussion, when $Q\rightarrow\infty$, ${\br}^{\rm{TR}}_{i}\rightarrow\bs_i$ and hence the transmit symbols can be perfectly recovered.

\vspace{2mm}
\subsection{TR-MRC with RDC windowing for OTFS in LTV channels}\label{sec:LTIsymb}
As opposed to LTI channels, time variations of LTV channels make it impossible to achieve temporal focusing with the same time-reversal filtering procedure as in Section~\ref{subsec:LTI}. Additionally, it is impossible to estimate the channel gains at each single time sample $\ell$, i.e., $h[k,\ell]$. Thus, we assume an approximately constant CIR over each OFDM symbol and consider the channel response to be known at the delay sample $D\in [0,M-1]$ in a given OFDM symbol $n$ where $D$ is the position of the isolated pilot symbol that is used for channel estimation, \cite{D}. Relying on the near constant CIR within each OFDM symbol, we can define the time-reversal filtering matrix for a given OFDM symbol $n$ as $[\bH^{\rm{TR}}_{n,q}]_{ab}={{{h}}}_{q}^* [ L-1-(a-b), Mn + D]$. Then, we substitute this time-reversal filter into (\ref{eq:rTR}) while considering the received signal in (\ref{eq:channel}) when the channel is LTV.

In contrast to the LTI channels, the correlation function of the LTV channels depends on the statistical properties of an additional random variable $\upsilon_{p,q}$. We note that the channel gains $\alpha_{p,q}$ and Doppler shifts $\upsilon_{p,q}$ at all the paths $p$ and all the BS antennas are independent with respect to one another. Therefore, $\expect \{ \alpha_{p,q} \Gamma(\upsilon_{p,q}) \} = \expect \{ \alpha_{p,q} \}\expect \{\Gamma(\upsilon_{p,q}) \}$ for any measurable function of $\upsilon_{p,q}$, $\Gamma(\upsilon_{p,q})$. Similar to our derivations in Section~\ref{subsec:LTI}, as $Q\rightarrow \infty$, the elements of the equivalent channel matrices in (\ref{eq:LTIrtr}), for LTV channels, tend to zero except for the diagonal elements of $\bG^{(n,n)}_{n}$, i.e.,
\vspace{0.1cm}
\begin{align}
    \label{eq:GiciLTV}
        [\bG^{(n,n)}_{n}]_{b,b}&= \frac{1}{Q}\sum^{Q-1}_{q=0}\sum^{L-1}_{k=0} {{{h}}}_q[k,Mn+b] {{{h}}}^*_q[k,Mn+D]\nonumber\\
        &= \frac{1}{Q}\sum^{Q-1}_{q=0}\sum^{L-1}_{k=0} \sum_{p=0}^{P-1}    e^{j2\pi (\upsilon_{p,q} (Mn+b-Mn-D)) T_{\rm{s}}} \nonumber \\ 
        & \hspace{5mm} \times \alpha_{p,q} \alpha_{p,q}^*  \delta [k-\ell_{\tau_{p,q}}] \nonumber\\ 
        &= \frac{1}{Q}\!\!\sum^{Q-1}_{q=0}\sum^{L-1}_{k=0} \sum_{p=0}^{P-1}\!\! | \alpha_{p,q}  |^2  e^{j2\pi \upsilon_{p,q} (b-D) T_{\rm{s}}} \delta [k-\ell_{\tau_{p,q}}].
\end{align}
\vspace{0.1cm}

From (\ref{eq:GiciLTV}), we observe that 
if the channel was LTI within each OFDM symbol, the same result as in the previous subsection would be achieved. However, in reality, the channel time variations over each OFDM symbol leaves a residual Doppler effect at each path that is observed on the exponential term $e^{j2\pi \upsilon_{p,q} (b-D) T_{\rm{s}}}$ in (\ref{eq:GiciLTV}). To gain an in-depth understanding of this effect on the performance of TR-MRC, we proceed with our asymptotic derivations that lead to our proposed RDC windowing technique.

As the number of BS antennas tends to infinity, the diagonal elements of $\bG^{(n,n)}_{n}$ in (\ref{eq:GiciLTV}), $[\bG^{(n,n)}_{n}]_{b,b}$ tend to $\sum^{P-1}_{p=0} \!\expect \{|\alpha_{p,q}  |^2\} \expect \{e^{j2\pi \upsilon_{p,q} (b-D) T_{\rm{s}}}\}$ where $\expect \{|\alpha_{p,q}  |^2\}\!\!\!=\!\!\!\rho(p)$. Thus, the term $\expect \{e^{j2\pi \upsilon_{p,q} (b-D) T_{\rm{s}}}\}$ needs to be calculated. Considering
the Jake's model, \cite{guo2019high}, for a given maximum Doppler frequency, $\upsilon_{\rm{max}}$, the Doppler frequency for path $p$ and antenna $q$ can be obtained as 
$\upsilon_{p,q}=\upsilon_{\rm{max}} {\rm{cos}}(\theta_{p,q})$,
where $\theta_{p,q}$ is uniformly distributed within the range $[-\pi, \pi]$, i.e. $\theta_{p,q} \sim \mathcal{U}(-\pi,\pi)$. 
Consequently, we have,
\begin{align}
\vspace{1mm}
    \label{eq:EVtheta}
        \expect \{ e^{j2\pi \upsilon_{p,q} (b-D) T_{\rm{s}}} \} &= \int^{\infty}_{-\infty} e^{j\beta {\rm{cos}}(\theta_{p,q}) (b-D)} f( \theta_{p,q} ) d\theta_{p,q} \nonumber\\
        &= {J_0}(\beta (b-D) ),
        \vspace{1mm}
\end{align}
where $\beta = 2\pi \upsilon_{\rm{max}}  T_{\rm{s}}$, $f(\theta_{p,q})=\frac{1}{2\pi}$ is the probability density function of $\theta_{p,q}$ and ${J}_0(\cdot)$ denotes the zero-order Bessel function of the first kind. This result shows that, in the asymptotic regime, the residual Doppler effect is reduced to a real valued attenuation function of the maximum Doppler frequency, given by
\be\label{eq:resultingexpectvalue}
    [\bG^{(n,n)}_{n}]_{b,b}=\sum^{P-1}_{p=0}\rho(p){J_0}(\beta (b-D))={J_0}(\beta (b-D)).
\ee

From (\ref{eq:resultingexpectvalue}), one may realize that, as $Q$ grows large, the residual Doppler effect can be easily corrected by dividing the TR-MRC output with the diagonal elements of $\bG^{(n,n)}_{n}$. Therefore, we propose an RDC windowing technique 
that can effectively compensate the residual Doppler effect. This window can be straightforwardly integrated into the window $\bW$ in (\ref{eq:rTR}) as $\bW=[\textbf{0}_{M\times L-1}, \bW_{J}]$, where
\be\label{eq:Wj}
    \bW_{J}={\rm{diag}}\{ [{J^{-1}_0}(  -D\beta),...,{J^{-1}_0}( (M-D-1)\beta) ]  \}.
\ee

Finally, the transmit data symbols can be recovered in the same fashion as in (\ref{eq:xtrLTI}). It is worth noting that our proposed RDC windowing technique in (\ref{eq:Wj}) brings substantial performance improvement for large relative velocities between the transmit and receive antennas. This relaxes the limitations of OTFS on the maximum Doppler frequency that it can tolerate. 

From implementation viewpoint, in light of the channel sparsity in delay-Doppler domain, it
can be misconstrued that implementation of our proposed technique in delay-Doppler domain is simpler than in delay-time. Taking a closer look, one may realize that delay-Doppler implementation requires one OTFS demodulator per antenna while delay-time implementation consists of only one OTFS demodulator irrespective of the number of BS antennas. Furthermore, based on the results of \cite{farhang2017low} on the channel impact, delay-Doppler implementation of (7) requires 2D convolution whereas its delay-time implementation requires 1D convolution operations. Hence, delay-time implementation of our proposed technique is a more practical choice than its delay-Doppler counterpart. 

\section{ Numerical Results}
\label{sec:numres}
In this section, we confirm the efficacy of our proposed TR-MRC with RDC windowing technique for OTFS without CP in massive MIMO channels through simulations. We consider the Extended Vehicular A (EVA) channel model, \cite{lte2009evolved}, and the Doppler shift is generated using Jake's model. In our simulation setup, we use the carrier frequency $f_{\rm c}=5.9$~GHz, subcarrier spacing $\Delta f=15$~kHz, $M=128$, $N=64$, $Q=200$ unless otherwise  stated. Transmission bandwidth is $B=330\Delta f=4.95$~MHz, 16-QAM modulation is deployed and we consider $5$ OTFS blocks per frame. For the relative speed of $V$~km/h, the maximum Doppler shift can be obtained as $\upsilon_{\rm{max}}=f_{\rm c}\frac{V}{3.6C}$ where $C \approx 3\times 10^8$~m/s is the speed of light. We assume perfect knowledge of the delay-time channel responses at all the BS antennas in the middle of the delay dimension when $D=\frac{M}{2}$.

\begin{figure}
    \centering
    \vspace{-2.9mm}
    \includegraphics[scale=0.63]{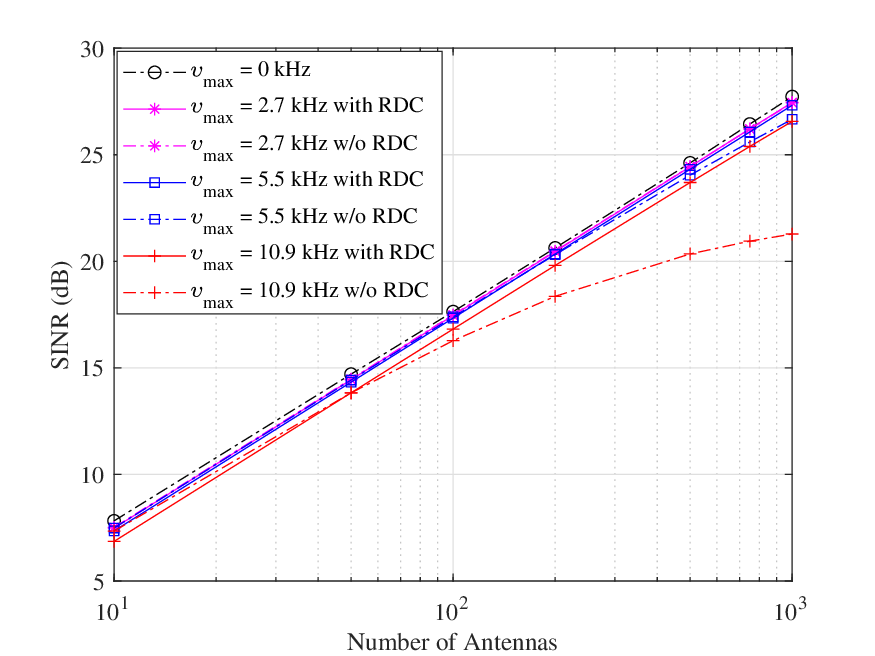}
    \caption{SINR performance as a function of the number of BS antennas.}
    \vspace{-0.4cm}
    \label{fig:SINRxAntBessel}
\end{figure}
\begin{figure}
    \centering
    \includegraphics[scale=0.63]{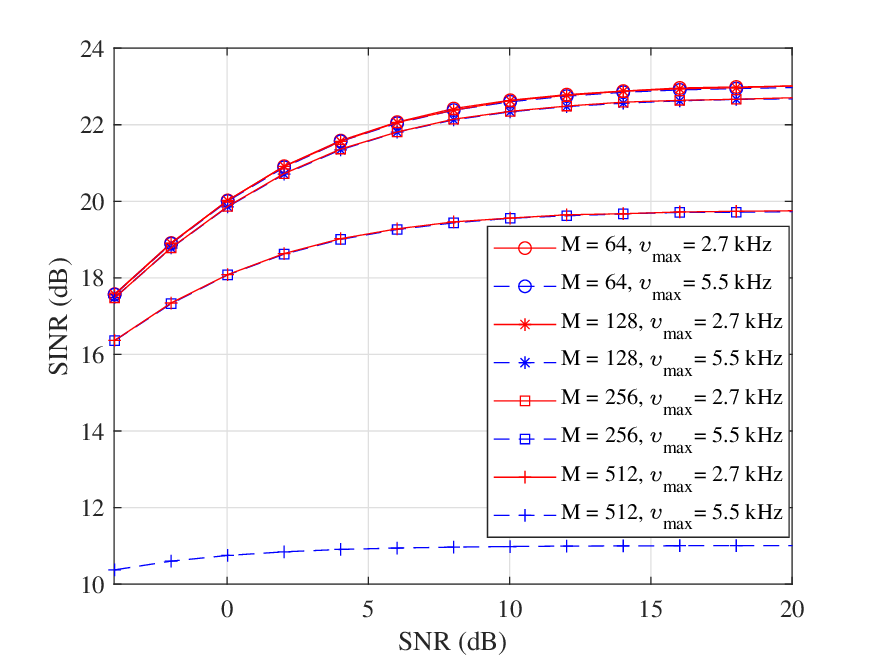}
    \vspace{-0.6cm}
    \caption{SINR performance versus input SNR for different symbol durations.}
    \label{fig:SINRxSNRwM}
    \vspace{-0.4cm}
\end{figure}

In Fig.~\ref{fig:SINRxAntBessel}, we compare the SINR performance of OTFS  as a function of the number of BS antennas for different maximum Doppler shifts with and without RDC windowing. Signal to noise ratio (SNR) at the input is $1$~dB. Based on our analytical results in Section~\ref{sec:TRinTD}, as the number of BS antennas grows large, TR-MRC can effectively average out the effect of ISI, IBI and time variations of the channel which is evident in Fig.~\ref{fig:SINRxAntBessel}. As the maximum Doppler shift increases, time variations of the channel within each OFDM symbol become more noticeable, and a residual Doppler effect remains after TR-MRC operation.
As shown in Section~\ref{sec:TRinTD}, this effect can be corrected by utilizing our proposed RDC windowing technique. The efficacy of our proposed technique is confirmed by the results in Fig.~\ref{fig:SINRxAntBessel} when the maximum Doppler shift is $\upsilon_{\rm{max}} = 10.9$~kHz which corresponds to the velocity  of $2000$~km/h at $f_{\rm c}=5.9$~GHz. It is worth noting that for $\upsilon_{\rm{max}} \leq 5.5$~kHz, corresponding to relative velocities up to $1000$~km/h, channel variations within each OFDM symbol is negligible. Consequently, the samples of the Bessel function in (\ref{eq:resultingexpectvalue}) take values close to $1$ and the RDC windowing brings a marginal amount of improvement. 

In Fig.~\ref{fig:SINRxSNRwM}, we present the SINR performance of OTFS with TR-MRC versus input SNR for different values of $\upsilon_{\rm{max}}$ and $M$ with a constant OTFS block size of $MN=8192$. From Fig.~\ref{fig:SINRxSNRwM} and equation (\ref{eq:resultingexpectvalue}), one may realize that every reduction/increase of the symbol duration, $M$, for a fixed $\upsilon_{\rm{max}}$, leads to the same result as if $\upsilon_{\rm{max}}$ was  reduced/increased with a fixed $M$ by the same proportion. This shows that the results for $\upsilon_{\rm{max}}=10.9$~kHz when $M=128$ would be the same for a smaller Doppler shift $\upsilon_{\rm{max}}=2.7$~kHz, corresponding to $V=500$~km/h, when $M=512$. Additionally, it is worth noting that in millimeter wave bands $\upsilon_{\rm{max}}=10.9$~kHz results from significantly lower speeds due to the high $f_{\rm c}$ values. However, when the Doppler shift is extremely large and/or symbol duration $M$ is very long, the Bessel function coefficients lead to a large amount of attenuation and thus the RDC window may lead to noise enhancement issues. Hence, it is the designer's choice to adjust the symbol duration $M$ and avoid or minimize the noise enhancement effects.

\begin{figure}
    \centering
    \vspace{-0.3cm}
    \includegraphics[scale=0.63]{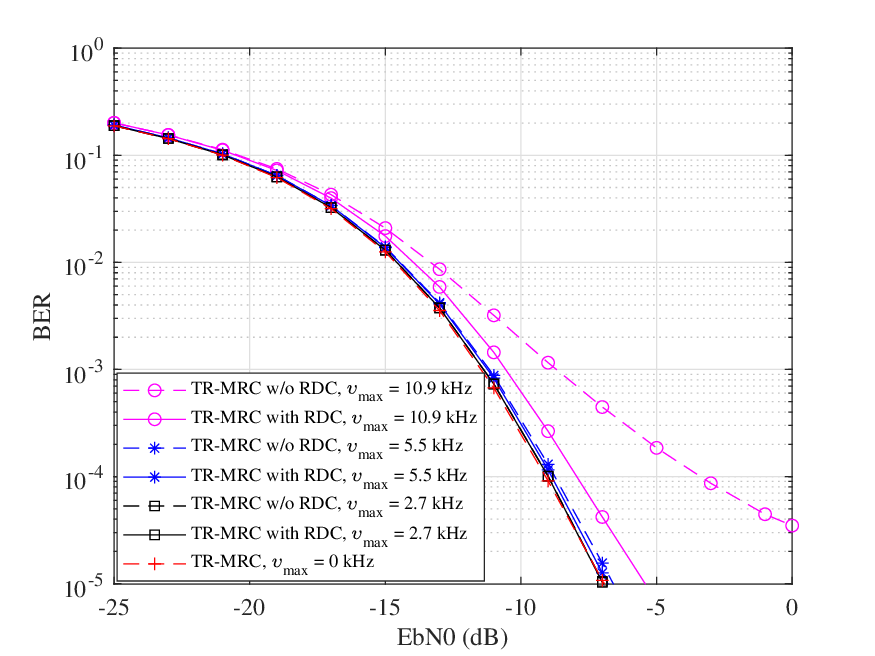}
    \vspace{-0.7cm}
    \caption{BER performance with RDC windowing.}
    \label{fig:BERvwB}
    \vspace{-0.4cm}
\end{figure}

Finally, in Fig.~\ref{fig:BERvwB}, we analyze the BER performance of our proposed TR-MRC technique for OTFS in the absence CP with and without RDC windowing. Fig.~\ref{fig:BERvwB} shows that for non-zero maximum Doppler shifts $\upsilon_{\rm{max}} \leq 5.5$~kHz, our proposed TR-MRC technique leads to about the same BER performance as the static scenario. This is while for $\upsilon_{\rm max}\approx11$~kHz, corresponding to the relative speed of $V=2000$~km/h, the proposed RDC windowing technique leads to the substantial performance improvement of over $5$~dB at high SNRs. This shows that our proposed technique brings the BER performance very close to that of the static scenario even for very large amounts of Doppler spread.

\section{~~Conclusion} \label{sec:conc}
In this paper, we developed a TR-MRC with RDC windowing technique as an enabler to completely remove the redundant CP from OTFS while breaking its Doppler limitations in massive MIMO systems. In particular, we asymptotically analyzed the performance of TR-MRC for OTFS without CP. We showed that both delay and Doppler spread of the wireless channel average out in the asymptotic regime when the maximum Doppler spread is within OTFS limits. However, when the maximum Doppler frequency goes beyond OTFS limits, a residual Doppler effect always remains. We analytically showed that this effect tends to a simple scaling of the received symbols in delay dimension as the number of BS antennas tends to infinity. The scaling coefficients are the samples of a Bessel function that is defined by the maximum Doppler shift. Based on this result, we proposed an RDC windowing technique that can break the Doppler limitations of OTFS and lead to a performance close to that of the LTI channels. The results in this paper can be extended to the multiuser case, where the multiuser interference is averaged out as the number of BS antennas tends to infinity. Further analysis on the multiuser scenario will be considered in our future work.

\bibliographystyle{IEEEtran}

\bibliography{main}

\end{document}